\documentclass[3p,times,procedia]{elsarticle}
\usepackage{color}

\usepackage{amssymb}
\usepackage{amsmath}

\makeatletter
\newcommand{\xRightarrow}[2][]{\ext@arrow 0359\Rightarrowfill@{#1}{#2}}
\makeatother

\usepackage[figuresright]{rotating}

\begin{document}

\begin{frontmatter}

\title{Collectivity in pp from resummed interference effects?}

\author{Boris Blok$^{1}$ and Urs Achim Wiedemann$^{2}$ }

\address{$^1$ Department of Physics, Technion -- Israel Institute of Technology,
 Haifa, Israel\\
$^2$ Theoretical Physics Department, CERN, CH-1211 Gen\`eve 23, Switzerland}

\begin{abstract}
Azimuthal asymmetries $v_n$ in the soft transverse momentum spectra of hadronic collisions can result as
a consequence of  quantum interference and color flow which translates spatial anisotropies into momentum anisotropies via 
multipole radiation patterns. Here, we analyze to what extent these effects result in signal strengths $v_n\lbrace 2s\rbrace$
that can persist in higher order $(2s)$ cumulants. 
In a simple model of soft multi-particle production with quantum interference effects in which $m$ particles are emitted from $N$
sources and in which interference contributions appear naturally ordered in inverse powers of the adjoint color trace, $1/(N_c^2-1)$,
we provide the first resummed calculation of all powers of $m^2/(N_c^2-1)$. This allows one to determine all higher order 
flow cumulants $v_n\lbrace 2s\rbrace$ with the same parametric accuracy. For a phenomenologically relevant range of $N$ sources
emitting $m$ particles,
we find that the even flow coefficients $v_n\lbrace 2s\rbrace$ decrease very mildly with increasing
cumulants. This provides a proof of principle that non-vanishing higher order cumulants $v_n\lbrace 2s\rbrace$ can persist
in systems that exhibit neither final state interactions nor phenomena related to high (saturated) initial parton densities. 
\end{abstract}

%% Preprint number  CERN-TH-2018-270

\end{frontmatter}
\vspace{0.5cm}

{\bf Introduction}.
Sizeable $n$-th harmonic coefficients $v_n\lbrace 2s\rbrace$ of azimuthal momentum asymmetries have been observed at the LHC in nucleus-nucleus (AA), proton-nucleus (pA) and proton-proton (pp) collisions~\cite{ALICE:2011ab,Abelev:2014mda,Khachatryan:2015waa,Aaboud:2017acw,Adare:2014keg,Adamczyk:2015xjc}. These asymmetries persist almost unattenuated if determined from higher order $(2s)$-particle cumulants, thus indicating
a collective mechanism that relates all particles produced in a given collision. The dynamical origin of this collectivity continues to be sought in competing and potentially contradicting 
pictures:

Explanations based on final state interactions are implicit e.g. in viscous fluid dynamic simulations~\cite{Romatschke:2017ejr}
and kinetic transport models \cite{Borghini:2010hy,Xu:2011fi,Uphoff:2014cba,Kurkela:2018qeb} of nuclear collisions. They
exploit that any interaction between the produced degrees of freedom implies transverse pressure gradients that 
translate spatial eccentricities in the overlap of the hadronic projectiles into momentum asymmetries $v_n\lbrace 2s\rbrace$. 
In AA collisions, jet quenching phenomena provide independent evidence that isotropizing final state interactions are indeed operational,  but
comparable evidence is missing in the smaller $pA$ and $pp$ collision systems. Moreover, in marked contrast to any final state explanation 
of flow anisotropies $v_n$ in pp collisions, the phenomenologically successful modeling of soft multi-particle production in modern multi-purpose pp event generators~\cite{Buckley:2011ms} are based on 
free-streaming partonic final state distributions supplemented by independent fragmentation into hadrons. Efforts to go beyond this picture are relatively recent, see e.g.~\cite{Fischer:2016zzs,Bierlich:2017vhg}. Therefore, two contradictory working hypothesis should be explored further: 
Either final state interactions are the cause for the measured $v_n\lbrace 2s\rbrace$ not only in AA but also in 
pp and pA hadronic collision systems -- this would invalidate the starting assumption of many underlying event models in pp collisions, and it would imply that quenching phenomena can be
found in pp and pA on some scale. Or there are dynamical mechanisms contributing to
the $v_n\lbrace 2s\rbrace$ that do not invoke final state interactions -- these would need to be taken into account in the no-final-state interaction baseline for analyzing $v_n\lbrace 2s\rbrace$ in AA collisions.
The present work makes a contribution towards exploring this second working hypothesis. 

Efforts to understand the measured $v_n$'s in terms of mechanisms operational in the incoming hadronic wave functions (a.k.a. initial state effects) have focussed so far mainly on parton saturation models, see e.g. Refs.~\cite{Levin:2011fb,Kovner:2011pe,Dumitru:2010iy,Dusling:2012iga,Kovchegov:2012nd} and subsequent work.
In these models, 
non-vanishing even second order cumulants $v_n\lbrace 2\rbrace$ result trivially from gluon emission of color dipoles. The calculation of higher order cumulants is, however, complicated,  
since the $S$-matrix is given in terms of eikonal Wilson lines $W$, and higher order 
cumulants involve target averages over a rapidly increasing number of $W$'s. By now, several ways of obtaining 
non-vanishing odd harmonics are identified, and there are calculations of fourth order and sixth order cumulants
~\cite{Dumitru:2014yza,Schenke:2015aqa,Altinoluk:2015uaa,Lappi:2015vta,McLerran:2015sva,Kovner:2016jfp}.
The most advanced model calculations~\cite{Dusling:2017aot,Mace:2018vwq}  provide phenomenologically satisfactory descriptions.
There is still debate on whether the modeling needed for this data comparison is quantitatively reliable~\cite{Nagle:2018ybc}, but the calculations
per se undoubtedly indicate that there can be contributions to $v_n\lbrace 2s\rbrace$ that do not invoke final state interactions.

{\bf Model of multi-gluon interference effects.} In Ref.~\cite{Blok:2017pui}, we have proposed
a simple model to study the effects of quantum interference and color flow on $v_n\lbrace 2s\rbrace$ without assuming large or saturated parton densities. The strong simplification of the model consists in neglecting a dynamically explicit formulation of the scattering process: all gluons in the incoming wave function are assumed to be freed in the scattering process with the same (possibly small) probability. 
The model pictures the incoming hadronic wavefunction as a collection of $N$ color sources in adjoint
representation distributed in transverse space according to a classical density $\rho({\bf y})$. On the amplitude level,
emission  of a gluon of color $a$ and momenum ${\bf k}_a$ from the $l$-th source is given by an eikonal factor $f({\bf k}_a)\, T^a\, e^{i {\bf k}_a.{\bf y}_l}$ where $T^a$ are generators of ${\rm SU}(3)$ in adjoint representation and ${\bf y}_l$ is the transverse position of the source. 
The cross section $ \textstyle\frac{d^m\hat\sigma}{d\Gamma_1d\Gamma_2 \cdots \, d\Gamma_{m}}$ for $m$-particle production from $N$ sources including interference effects is obtained by multiplying the sum of the $N^m$ emission amplitudes with its complex conjugate and summing (averaging) over all outgoing (incoming) colors,
  \begin{eqnarray}
	 && \frac{d^m\hat\sigma}{d\Gamma_1d\Gamma_2 \cdots \, d\Gamma_{m}}
  	\propto  N_c^m \left( N_c^2 - 1\right)^N  \left( \prod_{i=1}^m \left|\vec{f}({\bf k}_i)\right|^2 \right)  \label{eq1}
  		\\
  		 &&  \Big\lbrace  N^m + \sum_{d=1}^{m/2}
 	\left(F_{\rm corr}^{(2)}(N,m) \right)^d \frac{N^{m-2d}}{(N_c^2-1)^d} \sum_{(l_1m_1),(l_2m_2),...,(l_dm_d)} \sum_{(a_1b_1)(a_2b_2)\ldots (a_db_d)}
 	% \nonumber \\ && \qquad \times 
 	\prod_{j=1}^d	\left( 2^2\,  {\cos\left({\bf k}_{a_j}.\Delta {\bf y}_{l_jm_j} \right)}  { \cos\left( {\bf k}_{b_j}.\Delta {\bf y}_{l_jm_j}  \right)} \right)
  	+ \ldots  \Big\rbrace \, .
  	\nonumber 
 \end{eqnarray}
Here, $d\Gamma_i = k_i\, dk_i\, d\phi_i$ is the transverse phase space  of the $i$-th gluon,
$\Delta {\bf y}_{lm}  \equiv {\bf y}_{l} - {\bf y}_{m}$ is the transverse size of the source dipole $(l,m)$,
 and $\ldots$ stands for many other interference terms. We focus on the terms explicitly written. 
The first term in (\ref{eq1}) corresponds to {\it incoherent} emission of $m$ gluons that each link in the 
amplitude and complex conjugate amplitude to the same source (so-called diagonal gluons). For this first term, summing over initial and final color of each source leads to an adjoint color trace 
${\rm Tr}\left[1\right] = \left(N_c^2-1\right)$ and each gluon emission leads to a factor $\left|\vec{f}({\bf k}_i)\right|^2\,T^a T^a = N_c \left|\vec{f}({\bf k}_i)\right|^2$. 
Finally, as each of the $m$ gluons can be attached
 to one out of $N$ sources, there is an extra factor $N^m$. This explains all factors of the first term in eq.(\ref{eq1}). 
 As for the second term, we focus first on the contribution $d=1$ in the sum. This 
 contribution arises from squared amplitudes in which two gluons of color $a$ and $b$, emitted from two sources $l$ and $m$ interfer.  The resulting dipole interference term is 
 $\propto  2^2\,  {\cos\left({\bf k}_{a}.\Delta {\bf y}_{lm} \right)}  { \cos\left( {\bf k}_{b}.\Delta {\bf y}_{lm}  \right)}$, and it is suppressed by one power of the adjoint trace $(N_c^2-1)$ since
 the interfering gluons link the color flow between two sources $l$ and $m$~\cite{Blok:2017pui}. For such a contribution, $m-2$ gluons are diagonal thus leading to a factor $N^{m-2}$. As the sum $\sum_{(lm)}$ 
 goes over $N(N-1)/2$ dipole pairs, this second term is of the same $O\left(N^m\right)$ as the first one. Analogous arguments apply to contributions with $d>1$ dipoles in the sum of (\ref{eq1}), as long as none
 of the dipoles $(l_1,m_1)$, ..., $(l_d,m_d)$ shares a source with another dipole. 

In Ref.~\cite{Blok:2017pui}, we have shown that eq.(\ref{eq1}) gives rise to momentum asymmetries $v_2\lbrace 2\rbrace$ 
that coincide to leading $O\left(\textstyle\frac{1}{(N^2_c-1)}\right)$ with results of parton saturation models, and 
we have obtained explicit expressions for $v_2\lbrace 4\rbrace$ and $v_2\lbrace 6\rbrace$ in an expansion in powers of  
$\textstyle\frac{1}{(N^2_c-1)}$. However, corrections subleading in powers of $\textstyle\frac{1}{(N^2_c-1)}$ were found to 
be multiplied by factors $m^2$. This seems to narrow the range of validity of the naive $\textstyle\frac{1}{(N^2_c-1)}$-expansion 
to $m^2 < (N_c^2-1)$. Moreover, while $v_2\lbrace 4\rbrace$ and $v_2\lbrace 6\rbrace$ were found to be
non-vanishing, they have parametrically different $(N_c^2-1)$-dependencies in an expansion in powers of 
$\textstyle\frac{1}{(N^2_c-1)}$~\cite{Blok:2017pui}. Here, we show how resummation can overcome these limitations.
The main result reported in the present manuscript is a closed expression
that resums all leading contributions of order $\left(\textstyle\frac{m^2}{(N^2_c-1)}\right)^k$ and that yields for realistic multiplicity $m$ 
signal strengths $v_2\lbrace 2s\rbrace$ that are of the same parametric accuracy for all cumulants and whose numerical values
vary mildy with the order of the cumulant.

{\bf Calculating  azimuthal $(2s)$-particle correlation functions.} We want to calculate the correlation functions 
 \begin{equation}
 	K_{2s}^{(n)}(k_1,k_2,\cdots,k_{2s}) = {\cal N}
 \frac{ \int_\rho	\int d\phi_1 ...d\phi_{2s} \exp\left[in \left(\sum_{j=1}^s \phi_j - \sum_{j=s+1}^{2s} \phi_j \right) \right]\, \frac{d^{2s}N}{d\Gamma_1d\Gamma_2 \cdots \, d\Gamma_{2s}}}{ (2\pi)^{2s}
   \int_\rho \prod^{i=2s}_{i=1} \frac{dN}{d\Gamma_i} }\, ,
   \label{eq2}
 \end{equation}
 where $\int_\rho \ldots \equiv \int \left(\prod_i \rho({\bf y}_i)\, d{\bf y}_i \right) \ldots$ is the average over the transverse positions of the $N$ sources, and the standard $2s$-particle spectrum reads
 \begin{equation}
 	 	\frac{d^{2s}N}{d\Gamma_1d\Gamma_2 \cdots \, d\Gamma_{2s}} = \binom{m}{2s} \frac{d\hat\sigma}{\hat\sigma\, d\Gamma_1\, d\Gamma_2\cdots \, d\Gamma_{2s}}\, .
 	 	\label{eq3}
 \end{equation}
 In analogy to the experimental procedure of normalizing correlations by mixed event technique, the denominator in (\ref{eq2}) 
 is the product of one-particle multiplicity distributions. Its value 
$ \int d\Gamma_1 ...d\Gamma_{2s} \,\prod^{i=2s}_{i=1} \frac{dN}{d\Gamma_i} \equiv m^{2s}$ is the number of unordered choices 
of $2s$ particles from $2s$ different events that each contain $m$ particles. 
 The normalization in (\ref{eq3}) is chosen such that after integration over $2s$ one-particle phase spaces $d\Gamma_i$, eq.(\ref{eq3}) returns the number of possibilities $\binom{m}{2s}$ of 
 picking $2s$ out of $m$ particles.  This fixes the normalization ${\cal N} \equiv m^{2s} / \binom{m}{2s}$ in (\ref{eq2}) if one requires $K_{2s}^{(0)}=1$. We start by discussing the calculation of the numerator in (\ref{eq2}) that can be written as
\begin{equation}
	T^{(2s)} (\lbrace k_i\rbrace) = 
	 \int_\rho \left( \prod_{i=1}^{2s}  \int_0^{2\pi} d\phi_i\right) \,  \exp\left[in\left(\sum_{j=1}^s\phi_j - \sum_{j=s+1}^{2s} \phi_j\right) \right]
	\left(\int \prod_{b=2s+1}^m k_b\, dk_b\, d\phi_b \right)\,  \frac{d^{m}\hat\sigma}{d\Gamma_1d\Gamma_2 \cdots \, d\Gamma_{m}} \, .
	\label{eq4}
\end{equation}
 We first explain in which sense the terms written in (\ref{eq1}) are the parametrically dominant  interference contributions 
 for the calculation of (\ref{eq2}) and (\ref{eq4}), even though there are many other contributions that are not made explicit in (\ref{eq1}): 
 Diagrams in which a source carries only one gluon vertex in either amplitude or complex conjugate amplitude 
 vanish after color averaging over sources in the initial and final state, since ${\rm Tr}\left[ T^a\right] =0$.
 Diagrams that carry more than two vertices of off-diaginal gluons on a particular source are suppressed by powers of $N$. Therefore, leading contributions in the number $N$  of sources have exactly two off-diagonal gluon lines connected with each active source, or they emit diagonal gluons only. 
For diagrams with $m$ gluon lines, 
 $m-m_{\rm off}$ lines are diagonal and $m_{\rm off}$ are off-diagonal. In each such diagram, the 
$m_{\rm off}$ off-diagonal gluons can be grouped into a set of $l$ non-overlapping $n_i$-cycles~\footnote{To clarify the definition of $n$-cycle, we note: each dipole term $\propto  2^2\,  {\cos\left({\bf k}_{a}.\Delta {\bf y}_{lm} \right)}  { \cos\left( {\bf k}_{b}.\Delta {\bf y}_{lm}  \right)}$ can be viewed as a closed $2$-cycle with an off-diagonal gluon of momentum 
${\bf k}_a$ connecting 
sources $l \to m$ and the other off-diagonal gluon of momentum ${\bf k}_b$ connecting $m \to l$ and closing the cycle. Similarly,
three off-diagonal gluons that connect between sources $l_1\to l_2$,  $l_2\to l_3$ and  $l_3\to l_1$ form a closed
$3$-cycle, etc.} with $\sum_{i=1}^l n_i = m_{\rm off}$. 
 In general, each $n_i$-cycle links the color flow of $n_i$ previously independent sources and thus, compared to diagonal gluons, leads to a suppression of $n_i-1$ powers of the adjoint trace 
 ${\rm Tr}\left[1\right] = (N_c^2-1)$. Since $m_{\rm off} = \sum_{i=1}^l n_i$, the
 diagrams with $m_{\rm off}$ off-diagonal gluons that are of highest power in  $(N_c^2-1)$, are those that are organized in the
 maximal number $l$ of cycles. For $m_{\rm off}$ even, all such diagrams are therefore products of dipole emissions written explicitly in (\ref{eq1}).
 
To calculate the numerator $T^{(2)}$ of the two-particle correlation function (\ref{eq4}), 
all but $2$ off-diagonal momenta need to be integrated out. Since each phase space integration of an off-diagonal gluon comes
with a combinatorial factor $O(m)$, there is a multiplicative factor $m^{m_{\rm off} - 2}$ in $T^{(2)}$. On the other hand,
as explained above, contributions with $m_{\rm off}$ off-diagonal gluons are suppressed by order $1/(N_c^2-1)^{m_{\rm off}/2}$ or by
higher powers of $(N_c^2-1)$. The contributions to $T^{(2)}$ that are suppressed by the least powers of $(N_c^2-1)$ and that are 
enhanced  by the most powers of $m$ are therefore of order $ O\left(\textstyle\frac{m^{m_{\rm off} -2}}{(N_c^2 - 1)^{m_{\rm off}/2}}\right)$ for $m_{\rm off}$ even.
 The products of dipole terms written explicitly in (\ref{eq1}) are the only contributions to that order. 
Contributions with $m_{\rm off}$ odd contain at least one $n_i$-cycle with $n_i > 2$ and they will therefore give only
subleading contributions to $T^{(2)}$.  With an analogous line of argument, one checks that also for $s>1$, $T^{(2s)}$ receives all parametrically leading contributions from the terms
written explicitly in (\ref{eq1}). 

To write an analytically explicit expression for $T^{(2s)} (\lbrace k_i\rbrace)$, we introduce a short-hand for the phase space integral over a single dipole term 
 \begin{eqnarray}
 	A_2(k_a,k_b) &\equiv& \left|\vec{f}({\bf k}_a)\right|^2  \left|\vec{f}({\bf k}_b)\right|^2  \int_\rho
 	 \int  \frac{d\phi_a}{2\pi} \,  \int  \frac{d\phi_b}{2\pi}\, \exp\left[ i2 \left(\phi_a - \phi_b \right) \right]
 	 \cos\left({\bf k}_{a}.\Delta {\bf y}\right)  \cos\left( {\bf k}_{b}.\Delta {\bf y}  \right)\nonumber \\
 	 &=&  \left|\vec{f}({\bf k}_a)\right|^2  \left|\vec{f}({\bf k}_b)\right|^2\,   
 	  \int_\rho  J_2\left(k_{a} {\Delta y}\right)\, J_2\left(k_{b} {\Delta y}\right) 
 	  \approx \left|\vec{f}({\bf k}_a)\right|^2  \left|\vec{f}({\bf k}_b)\right|^2\,  \frac{1}{2} Bk_a^2 \, Bk_b^2\, .
	\label{eq5}
 \end{eqnarray}
Here, the last approximation is obtained for $k_a\, , k_b \ll 1/\sqrt{B}$ from a Gaussian source distribution $\rho({\bf y}) = 
\frac{1}{(4\pi B)^2} \exp\left[{\bf y}^2/{2B} \right]$ of spatial width $\sqrt{B}$. Such distributions $\rho({\bf y})$ arise naturally
in multi-parton interaction (MPI) models of the underlying event with $B \simeq (1 - 4) \,{\rm GeV}^{-2}$ fixed by
the measured MPI cross section in pp~\cite{Blok:2017pui,Paver:1982yp,Mekhfi:1985dv,Blok:2010ge}.
The approximation (\ref{eq5})  is of interest since it is valid in a physically relevant range 
$k_a\, , k_b \ll 1/\sqrt{B}$, where Bessel functions $J_2$ can be expanded for small arguments and final expressions
simplify considerably.

In calculating $T^{(2s)} (\lbrace k_i\rbrace)$ from eqs.~(\ref{eq1}) and (\ref{eq4}), one also finds factors that differ from 
$A_2$ by the absence of the  phase factors $e^{2i(\phi_{\bar a}-\phi_{\bar b})}$,
 \begin{eqnarray}
 	A_0 % &=& \int_\rho \int  k_a\, dk_a \, k_b\, dk_b\, \left|\vec{f}({\bf k}_a)\right|^2  \left|\vec{f}({\bf k}_b)\right|^2\,  \frac{d\phi_a}{2\pi} \, \frac{d\phi_b}{2\pi}\,  \cos\left({\bf k}_{a}.\Delta {\bf y}\right)  \cos\left( {\bf k}_{b}.\Delta {\bf y}  \right)\nonumber \\
 	 &=& \int_\rho  \int  k_{\bar a}\, dk_{\bar a}  \,  \left|\vec{f}({\bf k}_{\bar a})\right|^2
	\int  k_{\bar b}\, dk_{\bar b}\,  \left|\vec{f}({\bf k}_{\bar b})\right|^2   J_0\left(k_{\bar a} {\Delta y}\right)\, J_0\left(k_{\bar b} {\Delta y}\right)\, .
	\label{eq6}
 \end{eqnarray}
To write an explicit expression for $ T^{(2s)} (\lbrace k_i\rbrace) $ in terms of the shorthands (\ref{eq5}) and (\ref{eq6}) is now a straightforward but somewhat lengthy counting exercise:
 
We count $\binom{N}{2d}$ choices for attaching off-diagonal gluons to $2d$ out of $N$ sources. For these $2d$ sources,
there are $\textstyle \frac{(2d)!}{d! 2^d}$ different ways of combining them to dipoles. 

We next count the number of choices for picking $2(d-s)$ off-diagonal gluons such that a non-vanishing contribution arises
when integrating them out without phases in (\ref{eq4}). There are $\binom{m-2s}{ 2(d-s)}$ choices for choosing these gluons, 
and to distribute them amongst the  $d-s$ remaining dipoles, there are
$\binom{2(d-s)}{2} \binom{2(d-s-1}{2} \cdots \binom{4}{2} \binom{2}{2} = \frac{\left(2(d-s)\right)!}{2^{d-s}}$ possibilities. 

We next count the number of ways of assigning the $2s$ phases in (\ref{eq4}) to dipoles such that non-vanishing contributions arise.
First, since only terms of the form (\ref{eq5}) and (\ref{eq6}) can arise in the calculation of $T^{(2s)}(\lbrace k_i\rbrace) $,
the $2s$ phases appearing in $T^{(2s)} (\lbrace k_i\rbrace)$ must be combined to lie 
in $s$ out of the $d$ dipoles; for this there are  $\binom{d}{s}$ choices. Second, since exactly $s$ of the $2s$ phases in 
(\ref{eq4}) come with a plus sign, and since each non-vanishing contribution (\ref{eq5}) comes with one positive and
one negative phase, there are $s!$ choices to assign the positive phases times $s!$ choices to asign the negative one.
   
 Multiplying all the above mentioned factors yields
   \begin{eqnarray}
   \binom{N}{2d} \frac{(2d)!}{d! 2^d}
   \,
     \binom{d}{s}  s!^2 \binom{m-2s}{ 2(d-s)} \, \frac{\left(2(d-s)\right)!}{2^{d-s}}
% &=&   \binom{N}{2d} \frac{(2d)!}{d! 2^d}\,
%     \binom{d}{s}  s!^2
%     \frac{(m-2s)!}{2^{d-s}\, (m-2d)!} \nonumber \\
     &=&   \frac{N!}{(N-2d)!} \frac{s!}{(d-s)!}\,
     \frac{(m-2s)!}{(m-2d)!} \frac{1}{2^{2d-s}}\, .
     \label{eq7}
   \end{eqnarray}
In addition, one has $\binom{m}{2s}$ choices to pick $2s$ out of $m$ gluons in the calculation of (\ref{eq4}). We do not
include this factor in (\ref{eq7}), since it it is taken into account in the normalization of eq.~(\ref{eq2}). Combining these factors and denoting by
$U \equiv \int k\,  dk\, d\phi\, \left|\vec{f}({\bf k})\right|^2$ the phase space integrals over those $(m-2d)$ momenta that do not 
appear in the cosine-terms in (\ref{eq1}), one finds for the numerator of the $(2s)$-particle correlation function $K_{2s}^{(n)}(k_1,k_2,\cdots,k_{2s})$
\begin{eqnarray}
 T^{(2s)}(\lbrace k_i\rbrace) &=& % \binom{m}{2s}
 	N_c^m \left( N_c^2 - 1\right)^N  %\left( \frac{dN_{ch}}{d\eta}  \right)^{m}\, 
 	U^m\, N^{m}  \sum_{d=s}^{N/2} \left(F_{\rm corr}^{(2)}(N,m) \right)^d
% 	 \left( \frac{dN_{ch}}{d\eta}  \right)^{-2d}
	U^{-2d}
  		\nonumber \\
 	&& \times  \frac{1}{(N_c^2-1)^d}   \left( \frac{N!}{(N-2d)!\, N^{2d}}\right)\, 2^s\, \frac{s!}{(d-s)!}
 		\frac{(m-2s)!}{(m-2d)!}
	A_0^{d-s}\, {\rm perm}_s \left( A_2 \right)\, .
  	 \label{eq8}
\end{eqnarray}
Here, the permanent ${\rm perm}\left( A_2\right)$ of a matrix $A_s$ is defined in analogy to a determinant, but with the signs of all 
products of matrix elements positive irrespective of the signature of the permutation. For the $s\times s$=matrix 
defined by the entries $A_s(k_{a_i},k_{b_i})$, the permanent  ${\rm perm}\left( A_2\right) \equiv {\rm perm}_s\left( A_2\right)/s! $ 
allows one to write an explicit expression (\ref{eq8}) without taking recourse to the long wave-length limit 
$k_a, k_b \ll 1/\sqrt{B}$.  In the small-$k$-approximation of (\ref{eq5}), this simplifies to 
${\rm perm}_s\left( A_2\right) \Big\vert_{Bk_i^2 \ll 1} = \frac{1}{2^s}\prod_{i=1}^{2s} \left( \left|\vec{f}({\bf k}_i)\right|^2 B\, k_i^2 \right)$.
For the normalization of the correlation function $K_{2s}^{(n)}(k_1,k_2,\cdots,k_{2s})$ , we also need to determine $\hat\sigma$, 
which is the $s=0$ term in (\ref{eq8}),
\begin{eqnarray}
	\hat \sigma = {T}^{(0)} =  % \binom{m}{2s}
	 N_c^m \left( N_c^2 - 1\right)^N\, N^{m}  % \left( \frac{dN_{ch}}{d\eta}  \right)^{m} 
	 U^m\, \sum_{d=0}^{N/2} \, \left(F_{\rm corr}^{(2)}(N,m) \right)^d
%	 \left( \frac{dN_{ch}}{d\eta}  \right)^{-2d}
	U^{-2d}\, 
  		\frac{1}{(N_c^2-1)^d}   \frac{1}{d!} \left( \frac{N!}{(N-2d)!\, N^{2d}}\right)\,
 		\frac{m!}{(m-2d)!}
	A_0^{d} \, .
  	 \label{eq9}
\end{eqnarray}
The $(2s)$-particle correlation $K_{2s}^{(n)}(k_1,k_2,\cdots,k_{2s})$ in (\ref{eq2}) is then given by the ratio of (\ref{eq8}) and (\ref{eq9}),
 \begin{eqnarray}
 	K_{2s}^{(n)}(k_1,k_2,\cdots,k_{2s}) &=&  \frac{T^{(2s)}(\lbrace k_i\rbrace) \, % \left( \frac{dN_{ch}}{d\eta}  \right)^{2s} 
 	U^{2s} }{\hat\sigma\, 
 	\prod^{i=2s}_{i=1}  \left|\vec{f}({\bf k}_i)\right|^2 }  \nonumber \\
  &=& \frac{a^{-s}(m-2s)! \left(\textstyle\frac{N}{2} \right)!s!}{m! \left(\textstyle\frac{N}{2}-s\right)!} 
\frac{ _1 F_3\left(\lbrace s-\textstyle\frac{N}{2}\rbrace ;\lbrace \textstyle\frac{1}{2}, \textstyle\frac{1+m-N}{2}, \textstyle\frac{2+m-N}{2}\rbrace ;-\frac{N^2(N_c^2-1)}{16aF_2}\right)}{ 
 _1 F_3\left(\lbrace -\textstyle\frac{N}{2}\rbrace ;\lbrace \textstyle\frac{1}{2}, \textstyle\frac{1+m-N}{2}, \textstyle\frac{2+m-N}{2}\rbrace ;-\frac{N^2(N_c^2-1)}{16aF_2}\right)}\, 
 \prod_{i=1}^{2s} \left(B\, k_i^2 \right)\, .
\label{eq10}
 \end{eqnarray}
Eq.(\ref{eq10}) is the main result of this work.
 For the contribution to $K_{2s}^{(n)}$ that is leading in powers of $1/(N_c^2-1)^s$ and up to subleading orders $1/N$ in the number of sources, it resums correctly all corrections of order $\left(\textstyle\frac{m^2}{(N^2_c-1)}\right)^k$.
Remarkably, this resummation is given as an analytically known expression in terms of the generalized hypergeometric functions $_1F_3$. 
 Eq.(\ref{eq10}) is written in terms of 
the shorthand %$a \equiv \frac{A_0}{U^2} $
\begin{equation}
	a \equiv \frac{A_0}{U^2} = \frac{ \int_\rho  \int  k_{\bar a}\, dk_{\bar a}  \,  \left|\vec{f}({\bf k}_{\bar a})\right|^2
	\int  k_{\bar b}\, dk_{\bar b}\,  \left|\vec{f}({\bf k}_{\bar b})\right|^2   J_0\left(k_{\bar a} {\Delta y}\right)\, J_0\left(k_{\bar b} {\Delta y}\right)}{\left(\int k\,  dk\, d\phi\, \left|\vec{f}({\bf k})\right|^2\right)^2}\, ,
	\label{eq11}
\end{equation}
which
characterizes a dipole interference of gluons that carry transverse momentum $k_{\bar a}$ and $k_{\bar b}$ and that were produced
from sources separated by a transverse distance $\Delta y$. This term $a$ appears in $(2s)$-particle correlation functions in which the
particular momenta $k_{\bar a}$, $k_{\bar b}$ (and, a fortiori, the interference effects associated with these momenta) are integrated out. 
For $\Delta y \equiv \vert \Delta {\bf y}\vert = 0$, the interference effects in such terms are not geometrically suppressed,
and thus, $a$ is maximal when the Gaussian transverse width $\sqrt{B}$ of the source distribution $\rho(\Delta y)$ is negligible, 
$a\vert_{k_a, k_b \ll 1/\sqrt{B}} = 1$. In the opposite limit $\Delta y \to \infty$ of a widely extended source, $a$ in eq.~(\ref{eq11})
vanishes. As seen from the definition (\ref{eq11}), the value of $a$ depends on an interplay between the geometry 
and the shape of the spectrum $\propto  \left|\vec{f}({\bf k})\right|^2$
which determines to what extent the produced momenta can resolve a characteristic distance $\Delta y$. In the case of
pp collisions, the differences $\Delta y$ between different sources are on sub-femtometer scale (indicating that interference terms
between different sources are not negligible, $a > 0$), but some of the produced transverse momenta can resolve these distances
(indicating that interference effects are not maximal, $a < 1$).  To illustrate this generic situation, we choose $a= 0.1$
in the following. This is a typical value for $a$, if one uses in (\ref{eq11}) a transverse extension of $\rho$ consistent with 
constraints on the size of the proton wave funcation and a shape $ \left|\vec{f}({\bf k}_{\bar a})\right|^2$
consistent with the slope of  transverse momentum spectra.

{\bf Numerical results.}
From the normalized $(2s)$-particle correlations $K_{2s}^{(n)}(k_1,k_2,\cdots,k_{2s})$, we determine the higher order flow cumulants
\begin{eqnarray}
	&&v_n\lbrace 2\rbrace(k) =  \sqrt{ K_{2}^{(n)} }\, ,\qquad v_n\lbrace 4\rbrace(k) = \left( - \left(K_{4}^{(n)} - 2 {K_{2}^{(n)}}^2 \right)\right)^{1/4}\, ,
	\qquad v_n\lbrace 6\rbrace(k) = \left( \left(K_{6}^{(n)} - 9 K_{2}^{(n)} K_{4}^{(n)} + 12 {K_{2}^{(n)}}^3 \right)/4\right)^{1/6}\, ,\nonumber \\
	&&v_n\lbrace 8\rbrace(k) = \left( \left(K_{8}^{(n)} - 16 K_{2}^{(n)} K_{6}^{(n)} - 18 {K_{4}^{(n)}}^2 + 144 {K_{2}^{(n)}}^2K_{4}^{(n)}  - 144 {K_{2}^{(n)}}^4 \right)/33\right)^{1/8}\, ,
	\label{eq12}
\end{eqnarray}
where we follow the standard practice to evaluate the $K_{2s}^{(n)}$'s at $k_i=k$. In the limit $B\, k_i^2 \ll 1$ used to write eq.(\ref{eq10}), the $k$-dependence of all higher order cumulants is of the form 
$v_2\lbrace 2s\rbrace(k) =  v_2\lbrace 2s\rbrace(k=1/\sqrt{B})\, B\, k^2$. We note that the prefactor $v_2\lbrace 2s\rbrace(k=1/\sqrt{B})$ 
in this equation does not only characterize the curvature of $v_2\lbrace 2s\rbrace(k)$ at $k=0$, but it provides also  
a good proxy for the $k$-integrated value of $v_2\lbrace 2s\rbrace$. This can be seen from undoing the approximation in eq.~(\ref{eq5}) with the replacement $B\, k^2 \rightarrow 2\int_\rho \left( J_2(k\Delta y)\right)^2$. [We further note as an aside that with this replacement, 
one obtains a full $k$-dependence of $v_2\lbrace 2s\rbrace(k)$ that shares important commonalities 
with the experimentally observed one: it raises initially quadratically with $k$, it reaches a maximum at scale 
$k_{\rm max} \sim 1/\sqrt{B} = 1 - 2\, {\rm GeV}$ and it then falls off slowly with increasing $k$~\cite{Blok:2017pui}. ]
\begin{figure}[t]
 \vspace{-.3cm}
 \includegraphics[width=0.8\textwidth]{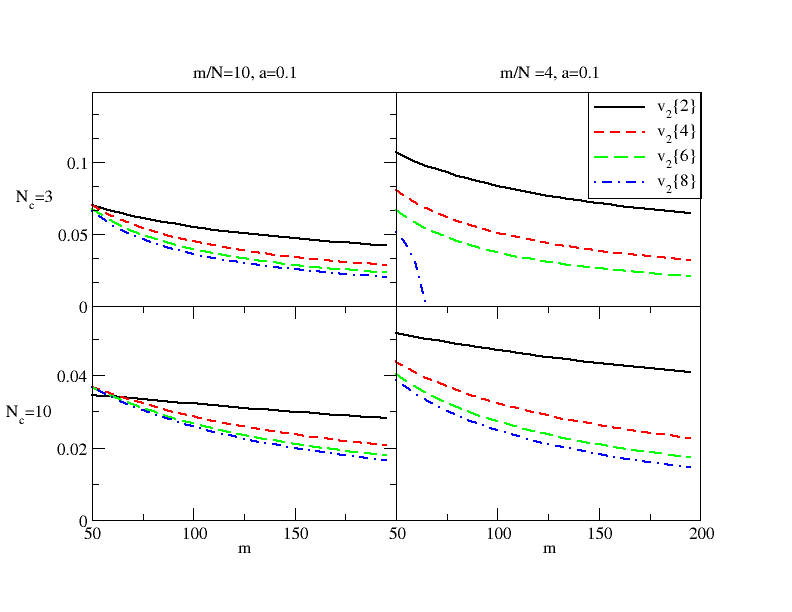} 
 \vspace{-1cm}
 \caption{ The elliptic flow cumulants $v_n\lbrace 2s\rbrace (k)$ of (\ref{eq12}), evaluated at momentum scale $B\, k^2=1$ from  $(2s)$-particle correlations functions (\ref{eq10}) in which all multiple dipole contributions to all orders $\left(\frac{m^2}{(N_c^2-1)}\right)^d$ are resummed.} 
 \label{fig1}
\end{figure}

In the following, we focus on the $s$-dependence of $v_2\lbrace 2s\rbrace$, and we do not explore further the $k$-dependence.
The very mild reduction of $v_2$-signals with increasingly higher order cumulants is regarded as a hallmark for collectivity in pp collisions. 
The main numerical result of this paper is the observation that for a suitable parameter range, a similar approximate persistence
of $v_2\lbrace 2s\rbrace$ with increasing $s$ can arise from a physical picture that invokes solely quantum interference and color 
correlation effects, see left hand side of Fig.~\ref{fig1}. Remarkably, the resummation of all powers of $\left( m^2/(N_c^2-1) \right)$ in eq.(\ref{eq10}) implies that all higher order cumulants are parametrically of the same order (in contrast to a corresponding calculation 
without resummation published in ~\cite{Blok:2017pui}), and it implies that within a certain parameter range, the numerical value of higher order
cumulants $v_2\lbrace 2s\rbrace$ decreases very mildly with $s$. 

As the above conclusions are limited to a certain parameter range, we now explain in some detail from where these limitations arise. 
To this end, we focus first on the $N$- and $a$-dependence of  $v_2\lbrace 2s\rbrace(k=1/\sqrt{B})$:
Eq.~(\ref{eq10}) is derived in the limit of many sources, when $1/N$ corrections are negligible. Thus, this derivation does not provide
insight into the finite $N$-dependence of $v_2\lbrace 2s\rbrace$. However, eq.~(\ref{eq10})
also contains an incomplete set of $1/N$ corrections; as numerical results are only meaningful in parameter ranges in which they are not
dominated by terms of order $1/N$, we have checked the stability of the results shown in Fig.~\ref{fig1} by setting in 
(\ref{eq10}) all terms of order $1/N$ explicitly to zero and repeating the calculation. For $a= 0.1$ and the ranges plotted in Fig.~\ref{fig1}, 
we confirm stability of the results against $1/N$ corrections. Also, while the absolute value $v_2\lbrace 2s\rbrace(k=1/\sqrt{B})$ 
changes when increasing $a > 0.1$, the relative $s$-dependence shows a very weak sensitivity to $a$, so that all the following
conclusions could be supported by a plot made for another value $a > 0.1$. 
However, for much smaller values $a < 0.1$, the numerical results start to become unstable since they start to be dominated
by the incomplete $1/N$ corrections. To explain this failure in the limit $a\to 0$, we note first that to leading order in $N$, 
eq.~(\ref{eq10}) is a resummation in powers of  $\left(\textstyle\frac{m^2 a}{(N^2_c-1)}\right)^k$ which reduces 
for $a\to 0$ to the unresummed $ v_2\lbrace 2s\rbrace$, which, as we know from  Ref.~\cite{Blok:2017pui}, is suppressed
by higher powers of $1/(N_c^2-1)$ which are not included in the calculation of (\ref{eq10}). Therefore, the leading $O(N^0)$ contribution
to $ v_2\lbrace 2s\rbrace$, $s \geq 4$ calculated here must vanish for $a \to 0$ (which we checked), and incomplete $1/N$ corrections can therefore dominate in this limit. This clarifies why the range of applicability of our calculation remains limited to $a>0.1$. 

We next turn to the multiplicity dependence of $ v_2\lbrace 2s\rbrace$. 
It is instructive to start this discussion with the academic limit of a system that emits a small number of gluons from a large
number of sources, $m \ll N$. In this case, the sum over the number of dipoles is limited in eqs.~(\ref{eq8}), (\ref{eq9}) by $m/2$
rather than by $N/2$. We have derived also for this limit analytical results for $K_{2s}^{(n)}$. We find that the value of 
higher order cumulants decreases rapidly with the order of the cumulant and, in this sense, the case $m \ll N$
is void of signs of collectivity. Only in the opposite case $m \gg N$ do we observe 
flow cumulants $ v_2\lbrace 2s\rbrace$ which decrease only mildly with increasing $s$, see the case for $m = 10\, N$ in Fig.~\ref{fig1}. 
For even larger values of $m/N$, differences between the higher order cumulants $ v_2\lbrace 2s\rbrace$ become even smaller
(data not shown). 
On the other hand, as the multiplicity $m$ moves closer to the number of sources $N$, first signs of
the break-down of collectivity are observed: for instance, for $m = 4\, N$, the eighth order cumulant
$v_2\lbrace 8 \rbrace^{8}$ in (\ref{eq11}) has the wrong sign in some range of $m$, and $v_2\lbrace 8 \rbrace$ in Fig.~\ref{fig1}
can therefore not be shown in that parameter range. As the derivation of (\ref{eq10}) is based on an expansion in powers of $1/(N_c^2-1)$,
this breakdown of collectivity can be pushed to smaller multiplicities in theoretical worlds with larger  $N_c$, see Fig.~\ref{fig1}. 
We thus find that collectivity (in the sense of a $v_2$-signal that is almost independent of the order of cumulant from which it is
calculated) will always be absent in the region $m \ll N$ and it will always be approximately realized in the region $m \gg N$.
We emphasize that in the present calculation,  gluons show azimuthal correlations irrespective of how far they 
are separated in longitudinal phase space~\cite{Blok:2017pui}. In this sense, $m$ is an event multiplicity and not a multiplicity  
per unit rapidity, and it is reasonable to assume that ultra-relativistic high-multiplicity pp collisions populate the range $m\gg N$.

In summary, we have demonstrated in a simple model that resummed quantum interference effects can lead to azimuthal flow
signals $v_n\lbrace 2s\rbrace$ that persist almost unattenuated in higher order cumulants. We caution
that the simple model studied here does not capture all observed flow phenomena. For instance, the $v_2$'s in Fig.~\ref{fig1}
are seen to decrease with increasing $m$ while the observed qualitative trend is seen in the data.\footnote{More precisely:
the particular $m$-dependence seen in Fig.~\ref{fig1} follows from a Glauber-like ansatz $m \propto N$. Varying this ansatz  
will yield a modified $m$-dependence. A signal $v_n\lbrace 2s\rbrace$ that increases with $m$ is therefore not excluded by
our study, but it is not a natural prediction of the framework explored here.}
 Our conclusion is therefore limited to the statement that the model calculation presented 
here provides a proof of principle
that quantum interference can contribute to flow-like multi-particle correlations even if both final state
rescattering effects and effects of parton saturation are absent.

We finally comment on the interpretation of our model calculation and on the 
relation of our results to calculations in parton saturation models. As we learnt from 
discussions with A. Kovner, the starting point of~\cite{Blok:2017pui} coincides with setting in CGC calculations (such as eq.(1) of Ref.~\cite{Altinoluk:2018ogz}) the target averages over Wilson lines to unity. Physically, Ref.~\cite{Blok:2017pui} can then either 
be interpreted as a model for final state gluon production based on the simplified assumption that all gluons in the initial
state are freed with the same (possibly small) probability. This is the point of view taken throughout this manuscript and in Ref.~\cite{Blok:2017pui}. Alternatively, the same calculation may be viewed as characterizing 
initial state effects: the calculation would indicate then that quantum interference and color flow in the in-state
can give rise to significant asymmetries in the intrinsic $k_T$-distribution of the incoming hadronic wave function. 
As one supplements this initial state interpretation with the assumption that the scattering process maps 
asymmetries in the intrinsic  $k_T$-distribution linearly to the final state, one regains the above-mentioned final state
interpretation. We close by repeating that the simplicity of the model studied here has allowed us to perform explicitly a resummation
of $O\left(m^2/(N_c^2-1) \right)$ that is required on physical grounds. Our calculation provides a proof of principle that momentum asymmetries that persist in higher order cumulants can arise from quantum interference and color flow alone.

\bibliographystyle{elsarticle-num}

\end{document}